\RequirePackage{fix-cm}
\documentclass[natbib,twocolumn,final]{svjour3}          
\usepackage{soul}
\usepackage{amssymb}
\usepackage{hyperref}
\usepackage{mathtools, cuted}
\usepackage{commath}
\usepackage{lineno}
\usepackage{siunitx}
\usepackage{graphicx,amssymb,amsmath}
\usepackage{txfonts}
\usepackage{natbib}
\usepackage{mathptmx}      
\usepackage{aps-bibstyle}  
\usepackage[first=0,last=9]{lcg}
 \journalname{to be inserted}
\newcommand{\aap}{{Astron. Astrophys.}}

\begin{document}
\title{The analysis of periodic orbits generated by Lagrangian solutions of
the restricted three-body problem with non-spherical primaries}
\author{ Amit Mittal\and
Md Sanam Suraj  \and
        Rajiv Aggarwal
}
\institute{Amit Mittal\at
Department of Mathematics,
ARSD College, University of Delhi, New Delhi-110021, Delhi,  India\\
 \email{\url{to.amitmittal@gmail.com}}
 \and
Md Sanam Suraj \at
    Department of Mathematics,
    Sri Aurobindo College, University of Delhi,  New Delhi-110017, Delhi, India\\
    \email{\url{mdsanamsuraj@gmail.com}}\\
    \email{\url{mdsanamsuraj@aurobindo.du.ac.in}}           
           \and
  Rajiv Aggarwal \at
  Department of Mathematics,
  Deshbandhu College, University of Delhi, New Delhi-110019, Delhi, India\\
              \email{\url{rajiv_agg1973@yahoo.com}}
 }
\date{Received:     date / Accepted:             date}
\maketitle
\begin{abstract}
The present paper deals with the  periodic orbits generated by Lagrangian solutions of the restricted three-body problem when both the primaries are oblate bodies. We have illustrated the periodic orbits for different values of $\mu, h,\sigma_1$ and $\sigma_2$ ($h$ is energy constant, $\mu$ mass ratio of the two primaries, $\sigma_1$ and $\sigma_2$ are oblateness factors). These orbits have been determined by giving displacements along the tangent and normal to the mobile coordinates as defined by Karimov and Sokolsky \cite{Kari}.  We have applied the predictor-corrector algorithm to construct the periodic orbits in an attempt  to unveil the effect of oblateness of the primaries by taking the fixed values of parameters $\mu, h, \sigma_1$ and $\sigma_2$.
\keywords{Restricted three-body problem\and Periodic orbit\and Oblateness\and Libration points}
\end{abstract}
\section{Introduction}
In the field of Celestial mechanics, the restricted problem of three bodies is most searched and fascinated problem for the most of astrophysicists.
The restricted three-body problem describes the motion of the infinitesimal test particle which moves in the gravitational field of two main primaries where they move in the circular or elliptic orbits around their common center of mass. The infinitesimal test particle does not influence the motion of the primaries. The available  literature on the restricted three-body problem unveiled the fact that the scientists and researchers have included various perturbation  terms  in the effective potential to obtain a realistic model. Indeed, this model can be used to study the solar system dynamics and kinematics  as well as in the study of stellar system. In particular, the model of the restricted three-body problem can be used in the space missions in an attempt to study the motion of the spacecraft in the Earth-Moon system.

 A plethora of research articles are available where the effect of various perturbations  due to radiations of primaries, the non-spheroid primaries, the effect of Coriolis and centrifugal forces and also the effect of variable mass is considered to study the dynamics  of the test particle, (e.g.,    \cite{AAG15, AMG15, abo15, AGV15, AAEA14, AGM14, AGM16, bha78, BC79, bha83},  \cite{R53},  \cite{SSA17, SS75, SS78, SI83,  SHC14,  SASC18, SHA14}).

The study of the families of periodic orbits in the restricted three-body problem has great importance in the field of Celestial mechanics. In the beginning of  last century, the families of the periodic orbits were studied by \cite{D09} and \cite{M20} firstly, but it was not complete.  In the few decades, a number of research paper were published where the families of periodic orbits were studied (e.g.,\cite{M58, M59}, \cite{G63},  \cite{DH65},  \cite{BG67},  \cite{MBM74}). Recently, a plethora of research papers are available where the  periodic orbits in the restricted four and five body problems are also investigated (e.g., \cite{ZP19}, \cite{PAA19}, \cite{KAE07}).

Ref. \cite{Char} and Ref. \cite{Plum} proved the existence of two families of small periodic motions near the Lagrangian solutions in the planar circular restricted three-body problem, with arbitrary values of the mass parameter. Ref. \cite{Ria} completed the analytical investigation of periodic motions.   The results on periodic motions of circular restricted three-body problem are presented in the famous book entitled "Theory of orbits" i.e., Ref. \cite{Sze}.  Ref. \cite{Dep} considered more results on periodic motions in their paper whereas Ref. \cite{Mar,  Mark} investigated the small periodic motions generated by Lagrangian solutions for all values of the mass parameter and for small values of energy constant  for which  the conditions of holomorphic integral theorem are valid. Ref. \cite{Had} has presented an exhaustive review of periodic solutions which are of interest to Dynamical Astronomy and their relation to actual systems. Ref. \cite{Kari} studied the periodic motions generated by Lagrangian solutions of the circular restricted three-body problem with the help of mobile co-ordinates by taking displacement along tangent and normal. Further, \cite{MAB09a} have extended their study by taking one of the primary as an oblate spheroid whereas the effects of radiation of the primary on the periodic orbits are investigated in \cite{MAB09b}.

Indeed, the celestial bodies are not spherical but  in general axis-symmetric bodies therefore, we thought of taking into account the shape of the bodies as well. The replacement of mass point by rigid-body is quite important because of its wide applications. Moreover, the re-entry of artificial satellite has shown the importance of periodic orbits.

That is why we have thought of studying, in this paper, to determine the periodic orbits generated by Lagrangian solutions of the restricted three-body problem when both the primaries are oblate bodies. We have generated the periodic orbits by giving the tangential and normal displacements to the mobile co-ordinates. In an attempt to disclose the effect of oblateness factor of both the primaries,  and varying value of energy constant, we have drawn the family of periodic orbits by fixing mass ratio of two primaries by applying the  predictor-corrector algorithm.

The structure of the paper is as follows: the description of the mathematical model and the equations of motion is explored in Sec. \ref{s:2}. The description of the normal and tangent variable are presented in the following Section \ref{s:3}. The methodology and the algorithm are described in the  Section \ref{s:4} for numerical simulations in an attempt to analyse the effect of oblateness of the primaries on the trajectories of the periodic orbits. The paper finally ends with Section \ref{s:5}, where the discussion and the conclusions are presented.
\section{The description of mathematical model and equations of motion}
\label{s:2}
The present system consists of two primaries $P_i, i=1,2$ and a test particle whose mass is infinitesimal in comparison of the mass of the primaries. We further assume that the primaries move in the  circular orbits under the mutual gravitational pull whereas the test particle moves in the combined gravitational pull of the primaries but not influencing their motion, in any way.

A rotating frame of reference is considered for the description of the dynamics of the test particle. In addition, the center of mass of the system is taken as the origin of the reference frame while the line joining the center of masses of the primaries are taken as $x-$axis. The dimensionless masses of the primaries are $m_1=\mu$ and $m_2=1-\mu$, where the mass parameter $\mu=m_1/(m_1+m_2)$ is same as Ref. \cite{Kari}. In addition, the centre of the primaries are on the horizontal axis with coordinates $(-\mu, 0)$ and  $(1-\mu, 0)$ whereas $(x,y)$ is the synodic rectangular dimensionless co-ordinates of the infinitesimal mass.

The equations of motion of the test particle with Lagrangian function $\mathfrak{L}$ are given by
\begin{subequations}
\begin{align}
\label{E:1a}
\frac{d}{dt}\frac{\partial \mathfrak{L}}{\partial \dot{x}}-\frac{\partial \mathfrak{L}}{\partial x}=&0,\\
\label{E:1b}
\frac{d}{dt}\frac{\partial \mathfrak{L}}{\partial \dot{y}}-\frac{\partial \mathfrak{L}}{\partial y}=&0,
\end{align}
\end{subequations}
where
\begin{align*}
\mathfrak{L}=&\frac{1}{2}\left(\dot{x}^2+\dot{y}^2\right)+n\left(x\dot{y}-\dot{x}y\right)+\frac{n^2}{2}\left(x^2+y^2\right)\nonumber\\
&+\frac{1-\mu}{r_1}+\frac{\mu}{r_2}+\sigma_1\frac{\mu}{2r_2^3}+\sigma_2\frac{1-\mu}{2r_1^3}+U,\nonumber\\
%
r_1&=\sqrt{\left(x+\mu\right)^2+y^2}, \\ r_2&=\sqrt{\left(x+\mu-1\right)^2+y^2},
\end{align*}
are the distances of the third body from the respective primaries $m_i, i=1,2$.

In the present paper, we have assumed that the shape of the primaries $P_i, i=1,2$ are not spherical but are oblate spheroid with oblateness coefficients $\sigma_i, i=1,2$, having masses $m_i$, while the mean motion $n$ is given by
\begin{align}\label{E:2}
 n=1+\frac{3}{4}\sigma_1+\frac{3}{4}\sigma_2,\quad \sigma_i<<1,
 \end{align}
 where,
\begin{align*}
\sigma_i&=\frac{a^2_i-c^2_i}{5R^2},  i=1,2, \\
R&=\text{is dimensional distance between primaries},\\
a_i, c_i&=\text{semi axes of the rigid body of mass} \ m_i.
\end{align*}
In addition, $U$  is a constant to be chosen in such a manner that the energy constant $h$ vanishes at the triangular libration point $L_4$.

\begin{figure}
\centering
\resizebox{\hsize}{!}{\includegraphics{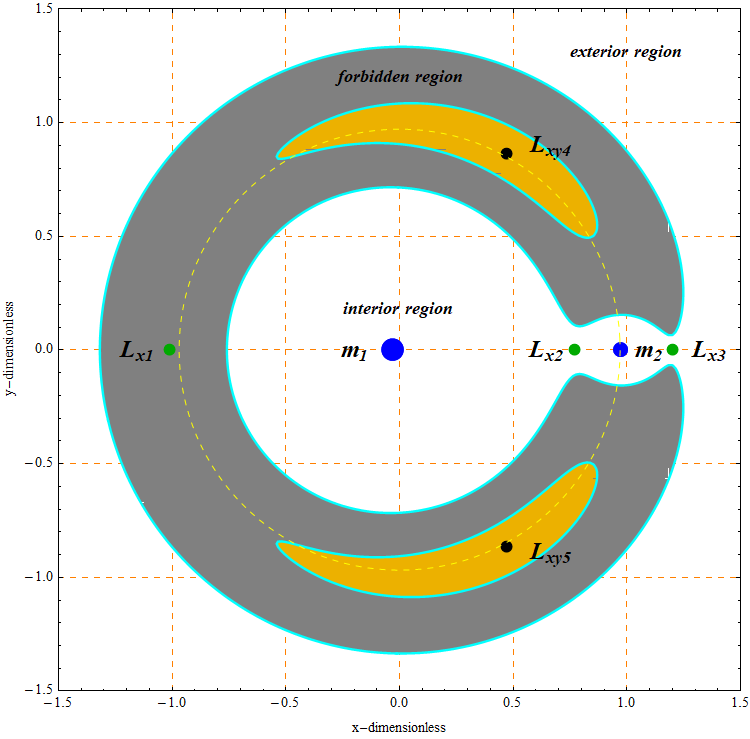}}
\caption{The zero velocity surfaces in which  the shaded regions show the forbidden regions of motion. The blue dots indicate
the location of the primaries, the black dots show the non-collinear  libration points, while green dots denote the collinear libration points.}
\label{Fig:01}
\end{figure}

The coordinates of $L_4$ are
\begin{align*}
x_{L_{4}}&=\frac{1}{2}\left(1-2\mu\right)+\frac{1}{2}\left(\sigma_2-\sigma_1\right), \\
y_{L_{4}}&=\frac{\sqrt{3}}{2}\Big\{1-\frac{1}{3}\left(\sigma_1+\sigma_2\right)\Big\},
\end{align*}
and
\begin{align*}
U=&-\frac{1}{2}(3-\mu+\mu^2)-\frac{\sigma_2}{4}(5-5\mu+3\mu^2)-\frac{\sigma_1}{4}(3\\
&-\mu+3\mu^2).
\end{align*}
Therefore, in dimensionless  synodic co-ordinate system, the equations of motion of the test particle can also be written as:
\begin{subequations}
\begin{align}
\label{E:3a}
\ddot{x}-2n\dot{y}&=W_{x},\\
\label{E:3b}
\ddot{y}+2n\dot{x}&=W_{y},
\end{align}
\end{subequations}
where
\begin{align*}
W&= \frac{n^2}{2}\left(x^2+y^2\right)+\frac{1-\mu}{r_1}+\frac{\mu}{r_2}+\sigma_1\frac{\mu}{2r_2^3}\\
&+\sigma_2\frac{1-\mu}{2r_1^3}+U+h.
\end{align*}
The corresponding Jacobi integral is read as:
\begin{align}\label{E:4}
C=&\frac{1}{2}\left(\dot{x}^2+\dot{y}^2\right)-\frac{n^2}{2}\left(x^2+y^2\right)-\frac{1-\mu}{r_1}-\frac{\mu}{r_2}\nonumber \\
&-\sigma_1\frac{\mu}{2r_2^3}-\sigma_2\frac{1-\mu}{2r_1^3}-U\equiv h,
\end{align}
which is the only Jacobian integral of motion,  exists for this dynamical system.
The regions of possible motion are illustrated in Fig. \ref{Fig:01}, and it is observed that as the energy constant decreases, the regions of possible motion increase. The shaded region shows the region where the motion of the test particle is forbidden  whereas the boundary of these forbidden regions shown in cyan color describes the zero velocity curves.
\section{Displacements along the normal and tangent variables}
\label{s:3}
We continue our study with the system of generalized coordinates $Q=(x,y)^T$ which depend upon four parameters $P=(\mu, h, \sigma_2, \sigma_1)^T$. The corresponding differential equations are given by system of Eqs. \ref{E:3a}, \ref{E:3b}, with Jacobi integral given by Eq. \ref{E:4}.  We determine the solutions of the Eqs. \ref{E:3a}, \ref{E:3b},  for which $C$ vanishes. If we consider the solutions of the Eqs. \ref{E:3a}, \ref{E:3b},   given by Eqs. \eqref{E:5} for some fixed parameters values $P=(\mu, h, \sigma_2, \sigma_1)^T$ then there may exist another solution given by Eqs. \eqref{E:6} with $P^*=(\mu^*, h^*, \sigma_2^*, \sigma_1^*)^T$ close to $P$. We have
\begin{equation}
\label{E:5}
\begin{cases}
x=x(t,\mu,h,\sigma_2,\sigma_1),\\
y=y(t,\mu,h,\sigma_2,\sigma_1),\\
\dot{x}=\dot{x}(t,\mu,h,\sigma_2,\sigma_1),\\
\dot{y}=\dot{y}(t,\mu,h,\sigma_2,\sigma_1),
 \end{cases}
 \end{equation}
and
\begin{equation}
\label{E:6}
\begin{cases}
x^*=x(t,\mu^*,h^*,\sigma_2^*,\sigma_1^*),\\
y^*=y(t,\mu^*,h^*,\sigma_2^*,\sigma_1^*),\\
\dot{x}^*=\dot{x}(t,\mu^*,h^*,\sigma_2^*,\sigma_1^*),\\
\dot{y}^*=\dot{y}(t,\mu^*,h^*,\sigma_2^*,\sigma_1^*).
\end{cases}
\end{equation}
The solutions in Eq. \eqref{E:6} will reduce to the solutions in Eq. \eqref{E:5} as $P^*\rightarrow P$.
We give the displacements
$\Delta P=P^*-P$ and $\xi=Q^*-Q$ to the parameters and the coordinates respectively as follows:
\begin{subequations}
\begin{eqnarray}
\label{E:7a}
\Delta \mu&=&\mu^*-\mu,\\
\label{E:7b}
\Delta h&=&h^*-h,\\
\label{E:7c}
\Delta \sigma_1&=&\sigma_1^*-\sigma_1,\\
\label{E:7d}
\Delta \sigma_2&=&\sigma_2^*-\sigma_2,\\
\label{E:7e}
\xi_1&=&x^*-x,\\
\label{E:7f}
\xi_2&=&y^*-y,
\end{eqnarray}
\end{subequations}
where $Q^*=(x^*,y^*)^T$ and $\xi=(\xi_1,\xi_2)^T$.

We consider that $\Delta P$ and $\xi$ are small quantities of the same order. Then, we have the following variational equations
\begin{subequations}
\begin{align}
\label{E:8a}
\ddot{\xi_1}=&W_{xx}\xi_1+W_{xy}\xi_2+2n\dot{\xi_2}+W_{x\mu}\Delta\mu+W_{xh}\Delta h\nonumber\\
&+W_{x\sigma_2}\Delta\sigma_2+W_{x\sigma_1}\Delta\sigma_1,\\
\label{E:8b}
\ddot{\xi_2}=&W_{yx}\xi_1+W_{yy}\xi_2-2n\dot{\xi_1}+W_{y\mu}\Delta\mu+W_{yh}\Delta h\nonumber\\
&+W_{y\sigma_2}\Delta\sigma_2+W_{y\sigma_1}\Delta\sigma_1,
\end{align}
\end{subequations}
with the integral, constructed from the Eq.  \eqref{E:4} by retaining the members of the first order only, given by
\begin{align}
\label{E:8c}
C=&\dot{x}\dot{\xi_1}+\dot{y}\dot{\xi_2}-W_x\xi_1-W_y\xi_2-W_\mu\Delta\mu-W_h\Delta h\nonumber\\
&-W_{\sigma_2}\Delta\sigma_2-W_{\sigma_1}\Delta\sigma_1.
\end{align}
We consider $V(t)=|\dot{Q}(t)|=\sqrt{\dot{x}^2+\dot{y}^2}$, as  momentary velocity on the orbit.

We assume that \eqref{E:6} is not corresponding to the equilibrium state, i.e., $V(t)\neq 0$ and we assume that $V(t)\neq 0$ on the whole orbit. Thus, $x$ and $y$ become the mobile coordinates. We will, now, use the mobile coordinate system to draw the periodic orbits by resolving one of the axes along the velocity vector $X=(\dot{x},\dot{y})^T$ and the other axis along the normal vector $Y=(-\dot{y},\dot{x})^T$. In the new coordinate system, we define the transition matrix $S$ as follows: $s(t)=X(t)/V(t)=$ the unit vector along the  tangent to the orbit is taken as the last column of the matrix $S$, whereas, the first column of $S$ is $r(t)=Y(t)/V(t)=$ the unit vector,  normal to the orbit which is orthogonal to the vector $s(t)$.

Therefore,, we have $S=\{r,s\}$, with dimension $2\times2$ as dim$(r)=2\times 1$, and dim$(s)=2\times 1$. We may verify that
\begin{subequations}
\begin{align}
\label{E:9a}
s^Ts &=1, \\
\label{E:9b}
r^Ts&=0,\\
\label{E:9c}
 \ S^{-1}s&=e=(0,1)^T.
\end{align}
\end{subequations}
 Finally, $S$ can be written as:
\begin{align*}
 &S=\frac{1}{V(t)}\left(\begin{array}{cc}
                  -\dot{y} & \dot{x}\\
                  \dot{x} & \dot{y}
                \end{array}\right),
\end{align*}
with its inverse as:
\begin{align*}
S^{-1}=S^T=\frac{1}{V(t)}\left(\begin{array}{cc}
                  -\dot{y} & \dot{x}\\
                  \dot{x} & \dot{y}
                \end{array}\right).
\end{align*}
We also define
\begin{align*}
&r^*=r^T=\frac{1}{V(t)}\left(-\dot{y}, \ \dot{x}\right), \ \textrm{as the first line of} \ S^{-1},\\
&s^*=s^T=\frac{1}{V(t)}\left(\dot{x}, \ \dot{y}\right), \ \textrm{as the last line of} \ S^{-1}.
\end{align*}
In the new coordinate system, we introduce
\begin{align*}
\alpha&=\left(\begin{array}{c}
               N \\
               M
             \end{array}
\right), \\
\textrm{dim}(N)&=1\times1, \text{and}\\
\textrm{dim}(M)&=1\times1,
\end{align*}
where $N$ is displacement along the normal to the orbit and $M$ is displacement along the tangent to the orbit.\\
Then, the new coordinates are given by the formulas
\begin{subequations}
\begin{align}
\label{E:10a}
\xi&=S\alpha=(r,s)\left(
                    \begin{array}{c}
                      N \\
                      M \\
                    \end{array}
                  \right)
=rN+sM,\\
\label{E:10b}
\alpha&=S^{-1}\xi, \\
\label{E:10c}
N&=r^*\xi,\\
\label{E:10d}
 M&=s^*\xi.
\end{align}
\end{subequations}
Now, differentiating Eq. \ref{E:10a}, we get
\begin{align*}
\dot{\xi}=\dot{S}\alpha+S\dot{\alpha}=\dot{r}N+r\dot{N}+\dot{s}M+s\dot{M},
\end{align*}
which can also be written as:
\begin{subequations}
\begin{align}
 \label{E:11a}
\xi_1=&\frac{1}{V}\left(-\dot{y}N+\dot{x}M\right),\\
\label{E:11b}
\xi_2=&\frac{1}{V}\left(\dot{x}N+\dot{y}M\right),\\
 \label{E:11c}
\dot{\xi_1}=&\frac{1}{V}\left(-\ddot{y}N+\ddot{x}M-\dot{y}\dot{N}+\dot{x}\dot{M}\right)-\frac{\dot{V}}{V^2}\Big(-\dot{y}N \nonumber\\
&+\dot{x}M\Big),\\
\label{E:11d}
\dot{\xi_2}=&\frac{1}{V}\left(\ddot{x}N+\ddot{y}M+\dot{x}\dot{N}+\dot{y}\dot{M}\right)-\frac{\dot{V}}{V^2}\left(\dot{x}N+\dot{y}M\right),
\end{align}
\end{subequations}
where
\begin{align*}
\dot{V}=\frac{\dot{x}\ddot{x}+\dot{y}\ddot{y}}{V}=\frac{W_x\dot{x}+W_y\dot{y}}{V}.
\end{align*}
Substituting these values into the integral (\ref{E:8c}), we have
\begin{align*}
C=&\frac{2}{V}\left(W_x\dot{y}-W_y\dot{x}+nV^2\right)N-\left(\dot{V}M-\dot{M}V\right)\\
&-W_\mu\Delta\mu-W_h\Delta h-W_{\sigma_2}\Delta\sigma_2-W_{\sigma_1}\Delta\sigma_1\equiv0,
\end{align*}
or
\begin{eqnarray}\label{E:12}
C&=&\frac{2W}{V^2}\left(\dot{M}V-M\dot{V}\right)+\frac{1}{V}\Big(W_x\dot{y}-W_y\dot{x}+\ddot{x}\dot{y}\nonumber\\
&&-\dot{x}\ddot{y}\Big)N+0\dot{N}-W_\mu\Delta\mu-W_h\Delta h-W_{\sigma_2}\Delta\sigma_2\nonumber\\
&&-W_{\sigma_1}\Delta\sigma_1\equiv0.
\end{eqnarray}
We may note that
\begin{align*}
&C=\frac{1}{2}V^2-W,\\
&\dot{C}=V\dot{V}-W_x\dot{x}-W_y\dot{y}\equiv0,
\end{align*}
we have $\frac{2W}{V^2}\equiv1$ for $C\equiv 0$.\\
The Eqn.\ref{E:12} can be solved for $\dot{M}$ as
\begin{eqnarray}
\label{E:13}
\dot{M}&=&\frac{M\dot{V}}{V}-\frac{1}{2W}\left(W_x\dot{y}-W_y\dot{x}+\ddot{x}\dot{y}
-\dot{x}\ddot{y}\right)N\nonumber\\
&&+\frac{1}{V}\left(W_\mu\Delta\mu+W_h\Delta h+W_{\sigma_2}\Delta\sigma_2+W_{\sigma_1}\Delta\sigma_1\right).\nonumber\\
&&
\end{eqnarray}
Now, using the equations
\begin{align*}
\ddot{x}&=2n\dot{y}+W_x,\nonumber\\
\ddot{y}&=-2n\dot{x}+W_y,\nonumber\\
\dddot{x}&=2n\ddot{y}+W_{xx}\dot{x}+W_{xy}\dot{y},\nonumber\\
\dddot{y}&=-2n\ddot{x}+W_{xy}\dot{x}+W_{yy}\dot{y},
\end{align*}
and substituted the values of $\dot{M}V-M\dot{V}$ from Eq.\ref{E:13}, the equations of motion Eqs. \ref{E:3a}-\ref{E:3b}, in the normal and tangent coordinates, can be written as:
\begin{align*}
S\ddot{\alpha}=\ddot{\xi}-2\dot{S}\dot{\alpha}-\ddot{S}\alpha,
\end{align*}
or
\begin{align*}
\ddot{\alpha}=S^{-1}\left(F_N N+F_{\dot{N}}\dot{N}+F_{\Delta P}\Delta P+\frac{\ddot{V}}{V}Ms\right),
\end{align*}
where
\begin{align*}
F_N=\left(\begin{array}{c}
            F_N^1 \\
            F_N^2
          \end{array}
\right),
F_{\dot{N}}=\left(\begin{array}{c}
            F_{\dot{N}}^1 \\
            F_{\dot{N}}^2
          \end{array}
\right),
\end{align*}
and the remaining symbols are available in the appendix.

Since $\alpha=\left(
                 \begin{array}{c}
                   N \\
                   M \\
                 \end{array}
               \right)
$ therefore, the equations of motion in normal and tangent co-ordinates can be written as
\begin{equation}\label{E:14}
\ddot{N}=r^{*}\left(F_{N}N+F_{\dot{N}}\dot{N}+F_{\Delta P}\Delta P\right),
\end{equation}
and
\begin{equation}\label{E:15}
\ddot{M}=\frac{\ddot{V}}{V}M+s^{*}\left(F_{N}N+F_{\dot{N}}\dot{N}+F_{\Delta P}\Delta P\right).
\end{equation}
The differential equation in the normal coordinate $N$ can also be written as:
\begin{align*}
\ddot{N}=&\frac{1}{V^{2}}\Biggl(W_{xx}\dot{y}^{2}-2W_{xy}\dot{x}\dot{y}+W_{yy}\dot{x}^{2}\\
&+\frac{1}{W}\left(-W_{x}\dot{y}+W_{y}\dot{x}\right)\left(-nV^{2}+\ddot{x}\dot{y}-\dot{x}\ddot{y}\right)\\
&-2n\left(\ddot{x}\dot{y}-\dot{x}\ddot{y}\right)
-\frac{1}{W}\left(\ddot{x}\dot{y}-\dot{x}\ddot{y}\right)\left(-nV^{2}+\ddot{x}\dot{y}-\dot{x}\ddot{y}\right)\\
&+\ddot{x}^{2}
+\ddot{y}^{2}-\dot{V}^{2}\Biggr)N+\frac{1}{V}\Biggl(\Biggl(-W_{x\mu}\dot{y}+W_{y\mu}\dot{x}\\
&+\frac{W_{\mu}}{W}
\left(-nV^{2}+\ddot{x}\dot{y}-\dot{x}\ddot{y}\right)\Biggr)\Delta \mu\\
&+\frac{W_{h}}{W}\left(-nV^{2}+\ddot{x}\dot{y}-\dot{x}\ddot{y}\right)\Delta h +\Biggl(-W_{x\sigma_{2}}\dot{y}\\
&+W_{y\sigma_{2}}\dot{x}+\frac{W_{\sigma_{2}}}{W}\left(-nV^{2}+\ddot{x}\dot{y}
-\dot{x}\ddot{y}\right)\Biggr)\Delta \sigma_{2}\\
&+\Biggl(-W_{x\sigma_{1}}\dot{y}+W_{y\sigma_{1}}\dot{x}+\frac{W_{\sigma_{1}}}{W}\left(-nV^{2}+\ddot{x}
\dot{y}-\dot{x}\ddot{y}\right)\Biggr)\Delta\sigma_{1}\Biggr).
\end{align*}

For the sake of simplicity,  we use the first order differential equation \ref{E:13} in $\dot{M}$ instead of using the second order differential equation \ref{E:15} in  $\ddot{M}$.
The matrix $S(t)$ can be taken as periodical and the Eqs. \eqref{E:14} and  \eqref{E:15} are linear differential equations with periodical coefficients at $\Delta P \rightarrow 0$ . We  have observed that the normal coordinate $N$ in Eq. \ref{E:14} is independent of the tangential coordinate $M$ and homogeneous.
\begin{figure*}
\centering
\resizebox{\hsize}{!}{\includegraphics{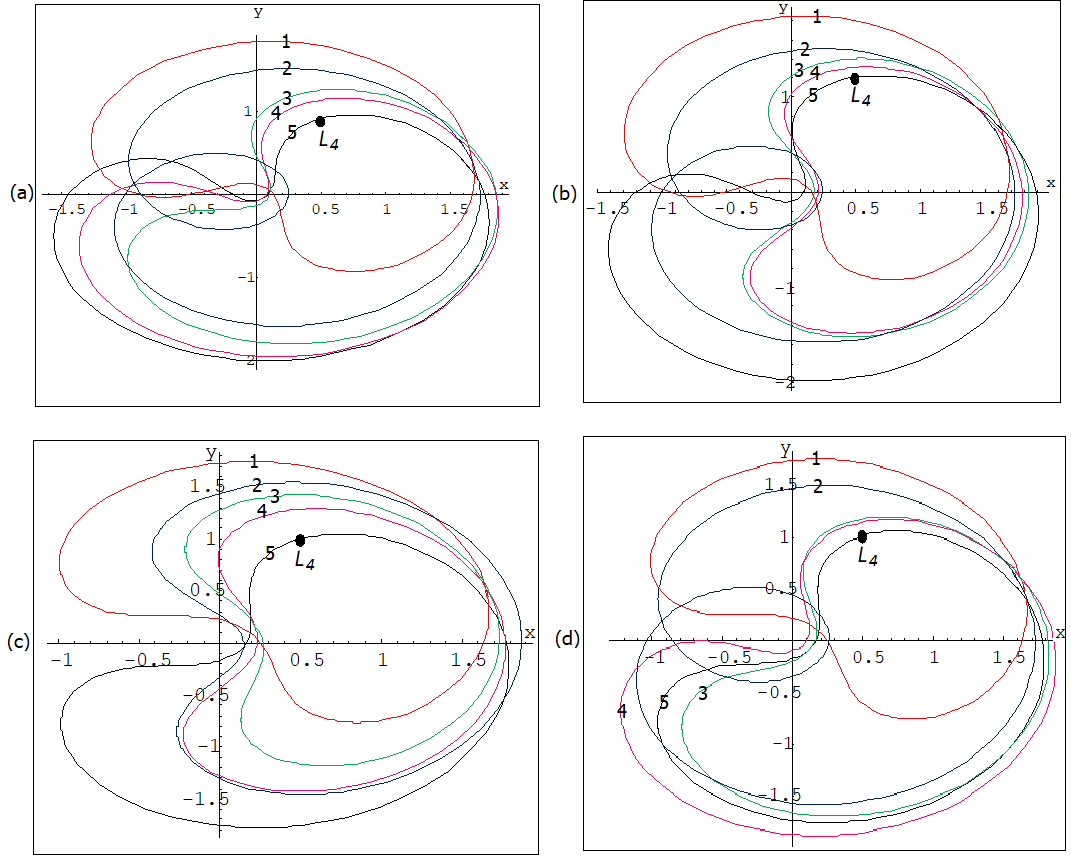}}
\caption{The periodic orbits: the color code are as follows: 1: red, 2: blue, 3: green, 4: rubin red, 5: black for $\mu = 0.001$, (a) $\sigma_1 = 0.0, \sigma_2 = 0.001$,  h1 = 0.05, h2 = 0.10, h3 = 0.15, h4 = 0.20, h5 = 0.328; (b) $\sigma_1 = 0.0001, \sigma_2 = 0.001$,  h1 = 0.40, h2 = 0.35, h3 = 0.30, h4 = 0.15, h5 = 0.1010;  (c) $ \sigma_1 = 0.001, \sigma_2 = 0.001$,  h1 = 0.30, h2 = 0.25, h3 = 0.20, h4 = 0.15, h5 = 0.1005;  (d) $ \sigma_1 = 0.001, \sigma_2 = 0.002$,  h1 =  0.30, h2 = 0.25, h3 = 0.20, h4 = 0.14, h5 = 0.10015.}
\label{Fig:1}
\end{figure*}
\section{The predictor-corrector method}
\label{s:4}
We develop an algorithm to find the periodic solution \eqref{E:6} in two stages. First predictor and then corrector. In the predictor part, we find linear displacements with respect to the parameters' increments for the initial conditions
\begin{align*}
x(0,\mu,h,\sigma_2,\sigma_1)&=x(T,\mu,h,\sigma_2,\sigma_1),\\
y(0,\mu,h,\sigma_2,\sigma_1)&=y(T,\mu,h,\sigma_2,\sigma_1),\\
\dot{x}(0,\mu,h,\sigma_2,\sigma_1)&=\dot{x}(T,\mu,h,\sigma_2,\sigma_1),\\
\dot{y}(0,\mu,h,\sigma_2,\sigma_1)&=\dot{y}(T,\mu,h,\sigma_2,\sigma_1),
\end{align*}
where $T$ is the  time period of periodic solution.

In the corrector part for the non-linear nature of parameters' increments, we use convergent iteration procedure to find the linear corrections (with respect to the parameters) in the initial conditions and the period.
\subsection{Predictor Part}\label{ss:4;1}
We introduce displacements $\xi$ by using the formula given in Eqs. \ref{E:7e}-\ref{E:7f}  and then normal and tangential displacements by using formula given in Eq. \ref{E:10a}. We use the Eq. \eqref{E:10a} to derive the Eqs. \eqref{E:13}, \eqref{E:14} and \eqref{E:15}. All the coefficients of $N$ and $M$ in these equations are periodic of period $T$.

The period $T^{*}$ of the desired solution can be written as $T^{*}=T+\tau$ , where $T = T(P), T^{*} = T(P^{*})$. Assuming $\xi$  and $\Delta P$ as first order small terms. By using the conditions of periodicity of the solutions for normal and tangential displacements, we have the following boundary conditions:
\begin{subequations}
\begin{eqnarray}
\label{eqn:16a}
N(0)&=&N(T), \ \dot{N}(0) = \dot{N}(T),\\
\label{eqn:16b}
M(0)&=&M(T)+V(0)\tau,\\
\label{eqn:16c}
\dot{M}(0)&=&\dot{M}(T)+\dot{V}(0)\tau.
\end{eqnarray}
\end{subequations}
The displacements in $N, M$ and $\tau$ can be written in linear combinations of varied parameters as follows:
\begin{subequations}
\begin{eqnarray}\label{E:18a}
N&=&N_{1}\Delta \mu + N_{2}\Delta h + N_{3}\Delta \sigma_{2} + N_{4}\Delta \sigma_{1},\\
\label{E:17a}
M&=&M_{1}\Delta \mu + M_{2}\Delta h + M_{3}\Delta \sigma_{2} + M_{4}\Delta \sigma_{1},\\
\label{E:17a}
\tau&=&\tau_{1}\Delta \mu + \tau_{2}\Delta h + \tau_{3}\Delta \sigma_{2} + \tau_{4}\Delta \sigma_{1}.
\end{eqnarray}
\end{subequations}
Using the independence of the parameter increments $\Delta P_{k}$ into the Eqs. \ref{E:12} and \ref{E:13} along with the boundary conditions given by Eqs. \ref{eqn:16a}-\ref{eqn:16c}, we have determined the value of $N_{k}, M_{k}$ and $\tau_{k}, (k = 1, 2, 3, 4)$.

Now, we determine the boundary problem for the normal displacement:
\begin{eqnarray}
&&\frac{dv_{k}}{dt}=\left(
              \begin{array}{cc}
                0 & I_{J} \\
                r^{*}F_{N} & r^{*}F_{\dot{N}} \\
              \end{array}
            \right)v_{k}+\left(
                           \begin{array}{c}
                             0 \\
                             r^{*}F_{\Delta P_{k}} \\
                           \end{array}
                         \right),\nonumber\\
&&v_{k}=\left(
        \begin{array}{c}
          N_{k} \\
          \dot{N}_{k} \\
        \end{array}
      \right),
v_{k}(0)=v_{k}(T), 
(k=1,2,\ldots, K),\nonumber\\
&&K=4 \ \textrm{and} \ J=1,\label{eqn:19}
\end{eqnarray}
where $F_{\Delta P_{k}}$ is the $k^{th}$ column of the matrix $F_{\Delta P}$ and $I_{J}$ is the unit matrix of dimension $J$.

The general solution of Eq. \ref{eqn:19} can be written as:
\begin{eqnarray}\label{eqn:20}
v_{k}(t)&=&Z(t)v_{k}(0)+v_{\Delta P_{k}}(t),
\end{eqnarray}
where $Z(t)$ is the matrix of fundamental solutions of homogeneous system with initial condition $Z(0)=I_{2J}, v_{k}(0)=$ initial conditions for $v_{k}(t)$ and $v_{\Delta P_{k}}(t)=$ a particular solution of inhomogeneous equations with zero initial conditions, i.e., $v_{\Delta P_{k}}(t)=0$.

Using Eqs. \eqref{eqn:19} and \eqref{eqn:20}, we obtain:
\begin{equation}\label{eqn:21}
v_{k}(0)=-\left(Z(T)-I_{2J}\right)^{-1}v_{\Delta P_{k}}(T).
\end{equation}
The Eqs. \eqref{eqn:20} together with Eqs. \eqref{eqn:21} give the solution of the boundary value problem given by Eqs. \ref{eqn:19}.

We determine displacement $\tau$ in the period by using the Eqn. \ref{E:12} in the form
\begin{equation}\label{eqn:22}
\dot{M}_{k}=\frac{\dot{V}}{V}M_{k}-\frac{V}{2W}\left(g_{v}v_{k}+g_{P_{k}}\right),
\end{equation}
where
\begin{align*}
g_{v}&=\left(g_{N} \ g_{\dot{N}}\right),\\
g_{N}&=\frac{1}{V}\left(W_{x}\dot{y}-W_{y}\dot{x}+\ddot{x}\dot{y}-\dot{x}\ddot{y}\right),\\
g_{\dot{N}}&=\left(\dot{x} \ \dot{y}\right)\frac{1}{V}\left(
                                                        \begin{array}{c}
                                                          -\dot{y} \\
                                                          \dot{x} \\
                                                        \end{array}
                                                      \right)=0,\\
g_{P_{k}}&=k^{th}  \ \textrm{element \ of \ row \ matrix},\\
g_{P}&=\left(-W_{\mu} \ -W_{h} \ -W_{\sigma_{2}} \ -W_{\sigma_{1}}\right).
\end{align*}
Then the general solution of Eqn. \eqref{eqn:22} is of the form:
\begin{equation}\label{eqn:23}
M_{k}(t)=\frac{V(t)}{V(0)}M_{k}(0)+\mu(t)v_{k}(0)+\mu_{\Delta P_{k}}(t),
\end{equation}
where row-vector $\mu(t)$ and $\mu_{\Delta P_{k}}(t)$ are the solutions of Cauchy problem \ref{E:26e} with \ref{E:26i} of system $(26)$ and $M_{k}(0)=0$.

Therefore, by using the periodical condition given in Eqn. \ref{eqn:16b}, we get
\begin{eqnarray}\label{eqn:24}
\tau_{k}&=&-\frac{1}{V(0)}M_{k}(T),\nonumber\\
&=&-\frac{1}{V(0)}\left(\mu(T)v_{k}(0)+\mu_{\Delta P_{k}}(T)\right).
\end{eqnarray}
Using $\tau_{k}$ by Eqn. \eqref{eqn:23} and  differentiating Eqn. \ref{eqn:22}, we obtain
\begin{equation}\label{eqn:25}
\dot{M}_{k}(0)=-\frac{V(0)}{2W(0)}\left(g_{v}(0)v_{k}(0)+g_{P}(0)\right).
\end{equation}
The new periodic motion can be determined by calculating the quantities in the right-hand sides of \eqref{eqn:20}, \eqref{eqn:23}, \eqref{eqn:24} and \eqref{eqn:25}. Therefore, we integrate the following differential equations from $t=0$ to $t=T$:
\begin{subequations}
\begin{eqnarray}
\label{E:26a}
&&\ddot{x}=2n\dot{y}+W_{x},\\
\label{E:26b}
&&\ddot{y}=-2n\dot{x}+W_{y},\\
\label{E:26c}
&&\dot{Z}_{j}=\left(
              \begin{array}{cc}
                0 & I_{J} \\
                r^{*}F_{N} & r^{*}F_{\dot{N}} \\
              \end{array}\right)Z_{j},\\
 \label{E:26d}
 && \ Z_{j}(0)=e_{j},  (j=1,2,\ldots 2J),\\
\label{E:26e}
&&\dot{\mu}_{j}=\frac{\dot{V}}{V}\mu_{j}-\frac{V}{2W}g_{v}Z_{j},\\
\label{E:26f}
&& \mu_{j}(0)=0,\\
\label{E:26g}
&&\dot{v}_{\Delta P_{k}}=\left(
              \begin{array}{cc}
                0 & I_{J} \\
                r^{*}F_{N} & r^{*}F_{\dot{N}} \\
              \end{array}
            \right)v_{\Delta P_{k}}+\left(
                           \begin{array}{c}
                             0 \\
                             r^{*}F_{\Delta P_{k}} \\
                           \end{array}
                         \right),\\
 \label{E:26h}
&&v_{\Delta P_{k}}(0)=0, (k=1,2,\ldots, K),\\
\label{E:26i}
&&\dot{\mu}_{\Delta P_{k}}=\frac{\dot{V}}{V}\mu_{\Delta P_{k}}-\frac{V}{2W}\left(g_{v}v_{\Delta P_{k}}+g_{P_{k}}\right),\\
\label{E:26j}
&&\mu_{\Delta P_{k}}(0)=0.
\end{eqnarray}
\end{subequations}
Here, $I_{2J}=\left(e_{1},\ldots, e_{2J}\right)$, $Z=\left(Z_{1},\ldots,Z_{2J}\right)$, $\mu=\left(\mu_{1},\ldots,\mu_{2J}\right)$, and the initial conditions $x(0), y(0), \dot{x}(0)$ and $\dot{y}(0)$ are known. The order of the above system is $2(I+J)+2J(2J+I)+(2J+1)K$, i.e., twenty two.

On solving the system of Eqs. \ref{E:26a}-\ref{E:26j}, we can find  $Z(T), v_{\Delta P_{k}}(T), \mu(T)$ and $\mu_{\Delta P_{k}}(T)$. Then, we determine  $N_{k}(0), \dot{N}_{k}(0), \dot{M}_{k}(0)$ and $\tau_{k}$ by using \eqref{eqn:21}, \eqref{eqn:25} and \eqref{eqn:24} respectively. We, further, calculate the values of $N(0)$, $ \dot{N}(0)$, $ M(0)$, $ \dot{M}(0)$ and $\tau$ from Eqs. \ref{E:18a}-\ref{E:17a}. The Eqs. \ref{E:11a}-\ref{E:11d} and (9a) give the values of $\xi(0)$ and $\dot{\xi}(0)$. Finally, we determine the initial conditions and period for new periodic solution with new parameter  $P^{*}=P+\Delta P$, using formulae:
\begin{eqnarray}\label{eqn:26}
x^{*}(0)&=& x(0)+\xi_{1}(0),\nonumber\\
y^{*}(0)&=& y(0)+\xi_{2}(0),\nonumber\\
\dot{x}^{*}(0)&=& a\left(\dot{x}(0)+\dot{\xi}_{1}(0)\right),\nonumber\\
\dot{y}^{*}(0)&= &a\left(\dot{y}(0)+\dot{\xi}_{2}(0)\right),\nonumber\\
T^{*}&=&T(P^{*})=T+\tau,
\end{eqnarray}
where
\begin{equation*}
a=\sqrt{\frac{2W(Q^{*}(0), P^{*})}{\dot{Q}^T(0)\dot{Q}^{*}(0)}},
\end{equation*}
is correction coefficient satisfying the energy integral given by Eq. \ref{E:4} if $a\equiv 1$. But, $C\neq 0$ as the predictor gives only approximate values of displacements in initial conditions.
\begin{figure}
\centering
\resizebox{\hsize}{!}{\includegraphics{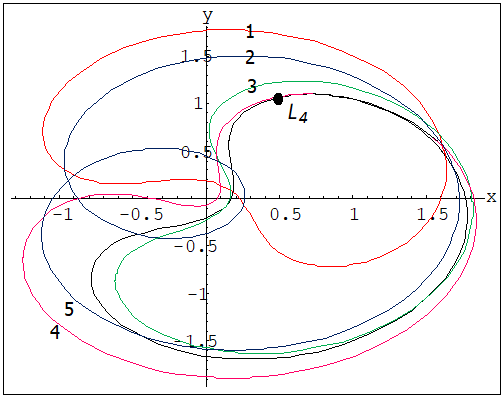}}
\caption{The periodic orbit for   $\mu=0.001, \sigma_1=0.002, \sigma_2=0.003$,  h1= 0.30, h2= 0.20, h3=0.15, h4=0.1025, h5=0.10010;}
\label{Fig:2}
\end{figure}
\subsection{Corrector Part}\label{ss:4;2}
This solution is periodic only approximately, i.e., the differences $x^{*}(T^{*})-x^{*}(0)$, $y^{*}(T^{*})-y^{*}(0)$, $\dot{x}^{*}(T^{*})-\dot{x}^{*}(0)$ and $\dot{y}^{*}(T^{*})-\dot{y}^{*}(0)$ are non-zero but small quantities of the second order with respect to $\Delta P$ in the previous step. To refine the initial conditions and the period, we use the corrector part only.

Let the solution \ref{E:5} obtained by the predictor part be non-periodic solution of the Eqs. \ref{E:3a}-\ref{E:3b} but in its close vicinity in the phase space there exists the periodic solution \ref{E:6} with same parameter values. Our goal is to find the periodic solution given by \ref{E:6} by taking the solution \ref{E:5} as the initial approximation.

We give the displacements by using the Eqs. \ref{E:7e}-\ref{E:7f} to determine displacements in initial conditions $\xi_{1}(0), \xi_{2}(0)$, $\dot{\xi}_{1}(0)$ and $\dot{\xi}_{2}(0)$ and the period $\tau$ at $\Delta P\equiv 0$.

The displacements $\xi_{1}(t)$ and $\xi_{2}(t)$ are satisfying the Eqs. \ref{E:8a}-\ref{E:8b} along with the Jacobi integral given by Eq. \ref{E:8c}. From the normal and tangential displacements given by Eqs. \ref{E:10c}- \ref{E:10d}, we obtain the differential equations \ref{E:12} and \ref{E:13}.

We consider the quantities
\begin{align*}
\Delta x&=x(T)-x(0)\neq 0,\\
\Delta y&=y(T)-y(0)\neq 0,\\
\Delta \dot{x}&=\dot{x}(T)-\dot{x}(0)\neq 0,\\
\Delta \dot{y}&=\dot{y}(T)-\dot{y}(0)\neq 0,
\end{align*}
as the small quantities of same order as that of  $\xi_{1}, \xi_{2}$, $\dot{\xi}_{1}, \dot{\xi}_{2}$ and $\tau$. Thus, the new boundary conditions for the Equations \ref{E:12} and \ref{E:13} are:
\begin{align}
N(0)&=N(T)+r^{*}(0)\Delta Q,\label{eqn:27}\\
\dot{N}(0)&=\dot{N}(T)+\dot{r}^{*}(0)\Delta Q+r^{*}(0)\Delta \dot{Q},\nonumber\\
M(0)&=M(T)+V(0)\tau+s^{*}(0)\Delta Q,\label{eqn:28}\\
\dot{M}(0)&=\dot{M}(T)+\dot{V}(0)\tau+\dot{s}^{*}(0)\Delta Q+s^{*}(0)\Delta \dot{Q},\nonumber
\end{align}
where $\Delta Q=\left(
                  \begin{array}{c}
                    \Delta x \\
                    \Delta y \\
                  \end{array}
                \right)
$
 and $\Delta \dot{Q}=\left(
                  \begin{array}{c}
                    \Delta \dot{x} \\
                    \Delta \dot{y} \\
                  \end{array}
                \right)
$.

The general solution of the Equation \ref{E:13} is determined in the same fashion as in the predictor-part  by using formula \ref{eqn:20} and rejecting high index $k$, i.e., $ v_{\Delta P_{k}}\equiv0$. Then by using boundary conditions given in Eqs. \ref{eqn:27} and \ref{eqn:28}, we get
\begin{align}\label{eqn:29}
v(0)=\left(
                  \begin{array}{c}
                    N(0) \\
                    \dot{N}(0) \\
                  \end{array}
                \right)=-\left(Z(T)-I_{2J}\right)^{-1}\left(
                                                        \begin{array}{c}
                                                          a_{1} \\
                                                          a_{2} \\
                                                        \end{array}
                                                      \right),
\end{align}
where
\begin{align*}
a_{1} =&\frac{1}{V(0)}\left(-\dot{y}(0)\Delta x+\dot{x}(0)\Delta y\right),\\
a_{2}=&\frac{1}{V(0)}\left(-\ddot{y}(0)\Delta x+\ddot{x}(0)\Delta y\right)-\frac{\dot{V}(0)}{V^{2}(0)}\Biggl(-\dot{y}(0)\Delta x\\
&+\dot{x}(0)\Delta y\Biggr)+\frac{1}{V(0)}\left(-\dot{y}(0)\Delta x+\dot{x}(0)\Delta y\right).
\end{align*}
The boundary problem for tangential displacement  can be found by substituting the value of normal displacement obtained by Eqs. \ref{eqn:29} in Eqn. \ref{E:12}. Now, we can find the general solution in the form of \eqref{eqn:23} by rejecting the index $k$ and assuming that $\mu_{\Delta P_{k}}\equiv0$.

if we set initial displacement along the orbit as zero, we get
\begin{eqnarray}
M(0)&=&0,\nonumber\\
\dot{M}(0)&=&-\frac{V(0)}{2W(0)}g_{v}(0)v(0)\label{eqn:30}.
\end{eqnarray}
Using the periodic conditions given by Eqs.  \ref{eqn:27} in \ref{eqn:28}, the displacement for the period is given by
\begin{align}
\tau&=-\frac{1}{V(0)}\left(\mu(T)v(0)+s^{*}(0)\Delta Q\right),\nonumber\\
&=-\frac{1}{V(0)}\left(\mu_{1}(T)N(0)+\mu_{2}(T)\dot{N}(0)\right)\nonumber\\
&+\frac{1}{V(0)}\left(\dot{x}(0)\Delta x+\dot{y}(0)\Delta y\right). \label{eqn:31}
\end{align}
In corrector part, we integrate \ref{E:26a}-\ref{E:26f} of system 26 from $t=0$ to $t=T$.
Hence, we calculate the values of $N(0)$, $ \dot{N}(0)$,
$ M(0)$, $ \dot{M}(0)$ and $\tau$ by using the Eqns. \eqref{eqn:29}, \eqref{eqn:30} and \eqref{eqn:31}.
Then, finally, we determine displacements $\xi(0)$ and $\dot{\xi}(0)$ by using \ref{E:8a}-\ref{E:8b}, and new initial conditions by using \eqref{eqn:26}. The above process is repeated again and again until a periodic orbit is drawn.
\section{The periodic orbits: discussion and conclusions}
\label{s:5}
We have determined periodic orbits in Figures 2 and 3  for fixed value of the mass parameter $\mu$, different values of oblateness parameters $\sigma_{1, 2}$ and varying values of the energy constant $h$. These orbits have been numbered $1, 2, 3, 4$ and $5$ corresponding to $h$ mentioned in each panel of the Figs. \ref{Fig:1} and \ref{Fig:2}. It is observed that the family in each case continues even if the orbit (number $5$) touches the point $L_4$.

In this paper, we have studied the periodic orbits associated with the mobile coordinates where $V(t)\neq0$ in the restricted three-body problem when both the primaries are oblate spheroid. We have introduced a new constant $U$ (similar to Karimov and Sokolsky \cite{Kari}) in the lagrangian function $\mathfrak{L}$ such that energy constant $h$ vanishes at the non-collinear libration point $L_{4, 5}$. We have drawn these periodic orbits by using the well known method: predictor-corrector. We have given the displacements to the mobile coordinates along the normal and the tangent directions to the orbit. We have plotted the five periodic orbits in a family by taking the fixed values of the mass parameter $\mu$, oblateness parameters $\sigma_{1, 2}$ and increasing values of the energy constant $h$. These periodic orbits are named as 1, 2, 3, 4, and 5 in each family. In order to draw these orbits, we have used the methodology adopted by Karimov and Sokolsky \cite{Kari}.

The most prominent observations on the periodic orbits as well as on the energy constant $h$ in the presence of the oblate primaries are as follows:

\begin{enumerate}
\item The energy constant $h$ decreases if we consider both the primaries as oblate bodies but it increases if we consider $\sigma_1=0$, i.e., in the absence of oblateness of the first primary.
\item With the increase in the oblateness parameters $\sigma_{1, 2}$, the energy constant $h$ decreases.
\item Each periodic orbit is non-symmetrical in the presence of oblateness of the primaries.
\item Family of periodic orbits do not terminate at the libration point $L_4$ rather they continued which is contrary  to the case of Karimov and Sokolsky (1989).
\end{enumerate}
We emphasize that this study is significantly different from others in the sense that most of the natural and artificial bodies moving in space instead of point masses, they are oblate bodies. Thus, our model is more realistic. Besides taking both the primaries as oblate
bodies, we have used mobile-coordinates by giving the displacements along the normal and the tangent to the orbit which has wider applications in space dynamics.
\section*{Appendix}
\begin{align*}
F_N^1=&\frac{\dot{x}W_{xy}-W_{xx}\dot{y}}{V}+\frac{1}{WV}\Big(-W_x \dot{y}+W_y\dot{x}\Big)\Big(n\dot{y}-\ddot{x}+\frac{\dot{V}}{V}\dot{x}\Big)\\
&+2n\frac{\ddot{x}}{V}-2n\frac{\dot{V}}{V^2}\dot{x}-\frac{1}{WV}\left(\ddot{x}\dot{y}-\dot{x}\ddot{y}\right)
\left(n\dot{y}-\ddot{x}+\frac{\dot{V}}{V}\dot{x}\right)\\
&+\frac{W_{xy}\dot{x}+W_{yy}\dot{y}-2n\ddot{x}}{V}-2\frac{\dot{V}}{V^2}\ddot{y}-\frac{\ddot{V}}{V^2}\ddot{y}
+2\frac{\dot{V}^2}{V^3}\ddot{y},
\end{align*}
\begin{align*}
F_N^2=&\frac{W_{yy}\dot{x}-W_{xy}\dot{y}}{V}+\frac{1}{WV}\left(-W_x \dot{y}+W_y\dot{x}\right)\left(-n\dot{x}-\ddot{y}+\frac{\dot{V}}{V}\dot{y}\right)\\
&+2n\frac{\ddot{y}}{V}-2n\frac{\dot{V}}{V^2}\dot{y}-\frac{1}{WV}\left(\ddot{x}\dot{y}-\dot{x}\ddot{y}\right)
\left(-n\dot{x}-\ddot{y}+\frac{\dot{V}}{V}\dot{y}\right)\\
&-\frac{W_{xx}\dot{x}+W_{xy}\dot{y}+2n\ddot{y}}{V}+2\frac{\dot{V}}{V^2}\ddot{x}+\frac{\ddot{V}}{V^2}\ddot{x}
-2\frac{\dot{V}^2}{V^3}\ddot{x},
\end{align*}
\begin{align*}
F_{\dot{N}}^1=&2\left(\frac{n\dot{x}}{V}+\frac{\ddot{y}}{V}-\frac{\dot{V}\dot{y}}{V^2}\right),\\
F_{\dot{N}}^2=&2\left(\frac{n\dot{y}}{V}-\frac{\ddot{x}}{V}+\frac{\dot{V}\dot{x}}{V^2}\right),\\
F_{\Delta P}=&\left(F_{\Delta\mu}^1 \ F_{\Delta h}^2 \ F_{\Delta\sigma_2}^3 \ F_{\Delta\sigma_1}^4\right),
\end{align*}
\begin{align*}
F_{\Delta\mu}^1=\left(\begin{array}{c}
                        W_{x\mu}+\frac{W_\mu}{W}\left(n\dot{y}+\frac{\dot{V}}{V}\dot{x}-\ddot{x}\right) \\
                        W_{y\mu}+\frac{W_\mu}{W}\left(-n\dot{x}+\frac{\dot{V}}{V}\dot{y}-\ddot{y}\right)
                      \end{array}
\right),
\end{align*}
\begin{align*}
F_{\Delta h}^2=\left(\begin{array}{c}
                        \frac{W_h}{W}\left(n\dot{y}+\frac{\dot{V}}{V}\dot{x}-\ddot{x}\right) \\
                        \frac{W_h}{W}\left(-n\dot{x}+\frac{\dot{V}}{V}\dot{y}-\ddot{y}\right)
                      \end{array}
\right),
\end{align*}
\begin{align*}
F_{\Delta\sigma_{2}}^3=\left(\begin{array}{c}
                        W_{x\sigma_{2}}+\frac{W_{\sigma_2}}{W}\left(n\dot{y}+\frac{\dot{V}}{V}\dot{x}-\ddot{x}\right) \\
                        W_{y\sigma_{2}}+\frac{W_{\sigma_2}}{W}\left(-n\dot{x}+\frac{\dot{V}}{V}\dot{y}-\ddot{y}\right)
                      \end{array}
\right),
\end{align*}
\begin{align*}
F_{\Delta\sigma_1}^4=\left(\begin{array}{c}
                        W_{x\sigma_1}+\frac{W_{\sigma_1}}{W}\left(n\dot{y}+\frac{\dot{V}}{V}\dot{x}-\ddot{x}\right) \\
                        W_{y\sigma_1}+\frac{W_{\sigma_1}}{W}\left(-n\dot{x}+\frac{\dot{V}}{V}\dot{y}-\ddot{y}\right)
                      \end{array}
\right).
\end{align*}
\section*{Acknowledgments}
\footnotesize
The authors are thankful to Center for Fundamental Research in Space dynamics and Celestial mechanics (CFRSC), New Delhi, India for providing research facilities.


\textbf{Compliance with Ethical Standards}
\begin{description}
  \item[-] Funding: The authors state that they have not received any research
grants.
  \item[-] Conflict of interest: The authors declare that they have no conflict of
interest.
\end{description}
\bibliographystyle{aps-nameyear}

\end{document}